# Multi-Agent Crisis Response systems - Design Requirements and Analysis of Current Systems

Khaled M. Khalil, M. Abdel-Aziz, Taymour T. Nazmy, Abdel-Badeeh M. Salem
Faculty of Computer and Information Science Ain shams University Cairo, Egypt
kmkmohamed@gmail.com, mhaziz67@gmail.com, ntaymoor@yahoo.com, absalem@asunet.shams.edu.eg

## ABSTRACT
Crisis response is a critical area of research, with encouraging progress in the past view yeas. The aim of the research is to contribute to building future crisis environment where software agents, robots, responders, crisis managers, and crisis organizations interact to provide advice, protection and aid. This paper discusses the crisis response domain requirements, and provides analysis of five crisis response systems namely: DrillSim [2], DEFACTO [15], ALADDIN [1], RoboCup Rescue [18], and FireGrid [3]. Analysis of systems includes systems' architecture and methodology. In addition, we identified features and limitations of systems based on crisis response domain requirements.

## Categories and Subject Descriptors
I.6 [**Simulation and Modeling**]: Applications, Model Validation and Analysis, Simulation Support Systems – Environments; I.2 [**Artificial Intelligence**]: General - Cognitive simulation, Problem Solving, Control Methods, and Search, Distributed Artificial Intelligence; J.7 [**Computers in Other Systems**]: Command and control, Real time.

## 1. INTRODUCTION
Crisis response can be viewed as iterative four interrelated sub-phases. The first is damage assessment, in which looses and their magnitudes are identified. The second is needs assessment, in which initially required response is identified. The third is prioritization of response measures, in which required response matches with available resources. If response demand is greater than the current available resources, decision makers must establish priorities or act for external resources. The fourth is actual response, in which crisis resources are deployed, and decisions are disseminated to responders and the population at large. During the four sub-phases, crisis response activities face challenge of reducing the influence of crises cause to society, the economy, and the lives of individuals and communities. Responders and crisis managers have to continuously adapt their behavior and make quick decisions to tackle unpredicted events. To effectively help responders and crisis managers carrying out their responsibilities and tasks; crisis response domain requirements should be identified to lead the development of crisis response systems.

The crisis response domain is characterized as a virtual environment of required distributed control, huge amount of data, uncertainty, ambiguity, multiple stakeholders with different objectives, and limited resources which continually vary [1]. In consequence of mentioned domain characteristics; crisis response systems require a multi-disciplinary system design approach. Crisis response systems design should include: (i) filtering and data fusion methods, (ii) decision-making and machine learning methods for determining actions in response to states, (iii) interaction mechanism to manage the interaction between multiple actors and to model collective behavior, and (iv) system architecture studies of different system organizations and information exchange topologies. Multi-agent Systems (MAS) have been advocated as the natural solution to the required design approach that necessitates some form of decentralized control within dynamic and uncertain environments. Research within a number of themes (Believe-Desire-Intension, and High-Level Architecture) in MAS is being pursued; considering different aspects of the interaction between autonomous agents and the decentralized system architectures. In addition, algorithms have been designed that allow agents to reconcile their constraints and preferences in order to maximize some global objective.

A number of crisis response systems have been developed based on multi-agents systems approach (such as: DrillSim [2], DEFACTO [15], ALADDIN [1], RoboCup Rescue [18], and FireGrid [3]) and more are being developed. A key aspect of such multi-agent based response will be agent-assisted crisis actors (first responder, managers, public) working together. Agents assist the crisis actor in planning, and determining resources to use.

Developed crisis response systems can be categorized according to system functionality into: (i) systems focus on handling specific crisis type, and (ii) systems focus on integrating sub-systems to build a framework of crisis response and management. These systems enable (i) more robust, interoperable, and priority-sensitive communications, (ii) better situational awareness and common operating picture, (iii) improved decision support and resource tracking, (iv) greater organizational agility, (v) better engagement of the public, and (vi) enhanced infrastructure survivability.

In this paper, we present analysis of current crisis response systems, taking in consideration crisis response domain requirements, agent-based design methodology, and systems' limitations. We conclude with the common limitations of current crisis response systems and a work we plan to carry out in the near future for proposed crisis response model for bird flu crisis response.

## 2. BACKGROUND

The origin of crisis response systems design approaches is early transferred from military systems; due to the similarity of operations characteristics. Crisis response operations characteristics are inherited from event surge requirements, managing risk that actions will be executed inappropriate by the fluidity of the situation, and managing the risk that actions will be rendered inappropriate because they were based on incomplete or inaccurate information. Crisis response may span a few hours to days or even months depending upon the magnitude and complexity of the event. Examples of developed military crisis response systems include: Analytical Conflict and Tactical Simulation (ACATS) for Weapons of mass destruction (WMD) crises, ALOHA for gas dispersion crises, Computer Assisted Protective Action Recommendation System (CAPARS) for chemical agent release, and HotSpot for chemical, biological or radiological agent dispersion [8]. Unfortunately, access to these military systems has been restricted since attack of 11$^{th}$ of September.

One of the design approaches is to mimic a crisis by conducting crisis drills over a sample region; incorporating information technologies in the process of response during the drill. Drills are expensive and scripted to given crisis situations. Also, large scale testing solutions are close to impossible to test via drills [2]. Another approach is to use simulation and modeling tools. Simulation and modeling tools allow creating what-if scenarios dynamically and determining the ability of the response to adapt to the changing crisis requirements. Actually, simulation and modeling approach has an extra benefit that reliable simulation model can be used for real-time support operations enhancing situational awareness and decision support [8].

Simulation and modeling systems for crisis response consist of a set of integrated tools which will differ based on the application they are designed for (Figure 1). Based on the definition of Integrated Emergency Response Framework (iERF), simulation and modeling tools include six types of tools. Planning tools are used for determination of impact of a crisis event, and/or aiding development of the response action plans and strategies. Vulnerability analysis tools are used for evaluation and assessment of response preparedness plans. Identification and detection tools are used for determining the possibility of the occurrence of crisis event Training tools are used for training response personnel for handling crisis events. Systems testing tools are used for testing of systems and equipments used for crisis response. Real-time response support tools are used for evaluation of the current/future impact of a crisis through real-time updates on the situation, and evaluation of alternative actions/strategies evaluations which are then used to direct the response actions on the ground.

The scope of the simulation tools can vary from national level modeling for large disaster events such as volcanic explosions, to modeling a city block for a scenario like a building explosion or fire. What have not been studied in as much detail are simulation tools that help understanding the system architecture and global response strategy influence on response operations.

## 3. SYSTEMS REQUIREMENTS

There exist huge problems in the current practice of crisis response operations matching responders and managers requirements. Response problems are projected as a combination of failure in communication, failure in technology, failure in methodology, failure of management, and finally failure of observation. Thus, the development of crisis response systems should be guided by the requirements of crisis response domain. There are many types of information that can be processed from heterogeneous information sources. Information should be filtered, summarized, and fused for crisis managers and responders. Crisis response systems should utilize planning, scheduling, task allocation, and resource management tools to help in formulating crisis management plans and tracking. Crisis organizations need to be cooperative; sharing information, and making appropriate decisions effectively.

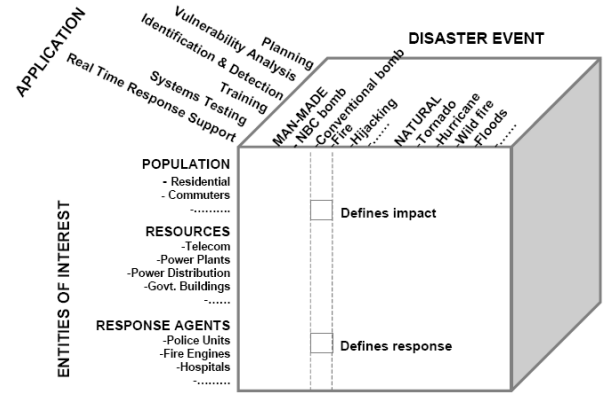

**Figure 1. Integrated Emergency Response Framework (iERF) proposed by NIST [8]**

### 3.1 Crisis response domain requirements
Crisis response domain requirements include:
a) Decentralized control towards response strategy.
b) Communication infrastructure is overloaded and is subject to damage.
c) Resources are limited and continually vary during response operations.
d) Uncertain and incomplete information from heterogeneous information sources which are required to be fused to support situational awareness and decision making processes.
e) Uncertainty and ambiguity of information about actors' actions results due to different actors' objectives and capabilities.
f) Adaptation of system components to environmental changes.
g) Adaptation of response plans to situation changes.
h) Integration of different tools.
i) Learning from experience.
j) Flexibility and rigidness.

### 3.2 Determining systems' aspects
Current crisis response systems are decentralized systems that inherit their capabilities and architecture from multi-agent based modeling methodology, and self-organized system aspects [21]. Self-organized system aspects describe macro (high-level) capabilities of crisis response systems. While, multi-agent systems aspects describe micro (low-level) capabilities. In tables (Table 1) and (Table 2), we list detailed aspects of self-organized system, and multi-agent systems matching crisis response domain requirements.

**Table 1. Crisis response domain requirements matching self-organized system aspects**

| Self-organized system aspects | | Crisis response domain requirements | | | | | | | | | |
|---|---|:---:|:---:|:---:|:---:|:---:|:---:|:---:|:---:|:---:|:---:|
| | | (a) | (b) | (c) | (d) | (e) | (f) | (g) | (h) | (i) | (j) |
| 1 | Design and build of systems that are fault-tolerant. | • | • | • | • | • | • | • | | • | • |
| 2 | Simplifying system maintenance by extending theme with some degree of plug-and-play functionality. | | • | • | | | • | | • | | • |
| 3 | Enable high level control of system at subsystem or system level. | • | • | | | | | | | | • |
| 4 | Extend the system functional scope by enable some degree of adaptation. | | • | • | • | • | • | • | | | • |
| 5 | Enable large collections of independent hardware/software components to coordinate their behaviors and strive for implicit collective goal. | • | • | • | | | • | • | • | | • |

**Table 2. Crisis response domain requirements matching multi-agent system aspects**

| Multi-agent system aspects | | Crisis response domain requirements | | | | | | | | | |
|---|---|:---:|:---:|:---:|:---:|:---:|:---:|:---:|:---:|:---:|:---:|
| | | (a) | (b) | (c) | (d) | (e) | (f) | (g) | (h) | (i) | (j) |
| 6 | Multi-agent system architecture (BDI, Cougaar, etc...) | • | • | • | | | • | | • | | • |
| 7 | Agents communication | • | • | | | | • | | • | | |
| 8 | Agent diversity | | | | • | • | | | • | | • |
| 9 | Institutional memory and knowledge management | | | | | | | | | • | |
| 10 | Intelligent actor behavior modeling (social interaction model) | | | | | | | • | | • | |
| 11 | Team formation (coalition formation) | • | • | • | | • | | • | | | |
| 12 | Coordination and task allocation | • | • | • | | | • | • | • | | |
| 13 | Collaboration and cooperation to achieve common goal | • | • | • | | | • | • | • | | • |
| 14 | Collaborative decision making | • | | | • | | • | • | | • | |
| 15 | Inference of local agent behavior, global agent behavior, and situation status | | | | | | • | • | • | | • | • |
| 16 | Adaptive planning | | • | • | • | • | • | • | | | • |
| 17 | Agent learning to refine rules and scenarios | | • | • | • | • | • | • | | • | |
| 18 | Security and safety of agents | • | | | | | | | • | | |
| 19 | Agent monitoring | | | | | | | | | • | |
| 20 | Agent behavior calibration | | | | | | | | | • | |

## 3.3 Interpretation

Concerning self-organizing methodology; aspects number 1 and 4 are supporting decentralized control, tolerance to resources failure, adaptation to unexpected changes, learning from experience and system flexibility. While, agent implementation aspects number 6, 12, 13, 16, and 17 are supporting adaptation to unexpected situations, handling uncertainties and ambiguity of information, integration of different tools, learning and system flexibility. By interpreting table columns; it is noticeable that self-organized systems and multi-agent systems methodology cover all crisis response domain requirements.

In what follows, we analyze crisis response systems based on previously mentioned self-organized and multi-agent systems aspects.

## 4. CRISIS RESPONSE SYSTEMS

**4.1 DrillSim [2], [16]:** DrillSim is an augmented reality user-centric simulation environment for testing IT solutions. The purpose of DrillSim is to play out a crisis response activity where agents might be either computer agents or real people playing diverse roles (first responders, crisis managers, experts, etc.). An activity in DrillSim occurs in a hybrid world that is composed of the simulated world generated by a multi-agent simulator and a real world captured by a smart space.

**Architecture:** DrillSim is a multi-agent simulation and modeling system. DrillSim is based on scalable architecture (O(100,000) agents), and is extended by plug-and-play capability. System components include I/O interfaces, simulation engine, data management module, database server for spatial data, and the Virtual Reality/Augmented Reality modules.

**Methodology:** Each agent has a role and a profile (age, cognitive abilities, health, and knowledge). Simulation scenarios are created by binding roles and profiles to agents. DrillSim has modeled agent behavior as a discrete process where agents alternate between sleep and awake states. Agents wake up and take some action every $t$ time units. For this purpose, an agent acquires awareness of the world around it, transforms the acquired data into information, and makes decisions based on this information using recurrent neural network. Then, based on the decisions, it

(re)generates a set of action plans using A* and object avoidance algorithms. Action plans dictate the actions the agent attempts before going to sleep again. Agents share and disseminate information based on their relationships (represented in a social network) via their own communication devices.

**Features:**
- System allows testing of IT solutions in the context of the simulated response activity to study the effectiveness of the solutions.
- System helps understanding the response activity.
- System integrates with other simulators (e.g., communication simulators, crisis simulators).
- Clear interfaces between information processing stages.
- Scenario based agent interaction (Q Language).
- Agents involved in information flow.
- Human can control and communicate with agents.
- Calibration of agent response models and metrics via running activity in simulated and real worlds.
- System keeps global log of every event, information exchange, and decision. In addition agents keep an individual consistent local log.
- Crisis real-time response support and training system.

**Limitations:**
- System architecture does not support fault-tolerance.
- System does not support high level control. Future actions are based on local agent behavior (operational level), permitting agents to execute undesirable action which leads to miss common operations goal.
- System does not support adaptive planning.
- The simulation engine includes the simulated geographic space, the evacuation scenario, and the agents. By such design approach, the switching to another implementation requires considerable reworks.
- Centralized simulation engine which leads to computational bottleneck on simulation server.
- Offline agent learning; agents need to learn about new roles, and information variables before scenario execution.
- Agent presents limited configurability in terms of decision, motion and health models, because their characteristics can be specified only through the hard-coded models.

## 4.2 Demonstrating Effective Flexible Agent Coordination of Teams through Omnipresence (DEFACTO) [15], [19]:

DEFACTO is a user centric system which incorporates 3D visualization omni-viewer, and human-interaction reasoning into a unique high fidelity system. Human-interaction allows responders to interact with the coordinating agent team in a complex environment, in which the responder can gain experience and draw valuable lessons that will be applicable in the real world.

**Architecture:** DEFACTO is a multi-agent simulation and modeling system based on Machinetta proxy architecture. Architecture of DEFACTO is scalable (O (10,000) agents) and flexible. DEFACTO consists of simulator, 3D omni-viewer, Machinetta proxy based teamwork infrastructure, and analysis tool to analyze the impact of teamwork interaction strategies.

**Methodology:** DEFACTO has modeled agent in proxy team formation (Machinetta). Machinetta proxies are responsible for transfer-of-control over a decision, managing local team beliefs, communication between proxies, communication between proxy and a team member, coordination, and task allocation for the team. Each proxy provides all transfer-of-control strategy options. One of strategy options is selected based on current situation and agent role. An optimal transfer-of-control strategy balances the risk of high quality decision made by human against the risk of costs incurred due to a delay in getting the decision from agent. Each team implements team-oriented plans which describe joint activities to be performed. Joint activities may include duplicate or conflicting tasks; hence Machinetta includes conflict resolution algorithms to remove conflicts.

**Features:**
- Improved situational awareness via interactive omni-viewer.
- Improved team performance through flexible human-agent interaction strategies.
- System allows transferring control from human to agents' team using team-level strategies.
- Conflict resolution algorithms.
- System divides global goal (strategic level) into sub-goals (tactical level) represented as joint intentions; which are executed via agent team members (operational level).
- Coordination, collaboration, and task allocation between agent team members.
- Calibration of human-agent transfer-of-control strategies
- Crisis real-time response support and training system

**Limitations:**
- System does not support fault-tolerance.
- System does not support plug-and-play capability.
- Building 3D model for omni-viewer can require months or even years of manual modeling efforts.
- Agents need high bandwidth communication channels to communicate.
- Agent presents limited configurability in terms of agent profile and scenarios.
- System does not support adaptive planning.
- System does not provide learning from experience strategies.

## 4.3 Autonomous Learning Agents for Decentralized Data and Information Systems (ALADDIN) [1], [5]:

ALADDIN is a user centric system, which aims to model, design, and build decentralized systems that can bring together information from variety of heterogeneous sources in order to take informed action. For that goal, ALADDIN is considering different aspects such as data fusion, decision making, machine learning, and system architecture.

**Architecture:** ALADDIN architecture is based on High Level Architecture (HLA) standard. ALADDIN organizes the simulator software in four different layers: (v) Simulation Model Layer, (iii) Simulation Components Layer, (ii) Discrete Event Simulation Layer, and (i) Distributed Discrete Event Simulation Layer. Simulation Model Layer is the layer where the simulation model is defined through the declaration of the agents involved in the simulation. ALADDIN is scalable and reusable system working through decentralized simulation framework.

**Methodology:** System is composed of autonomous, reactive, and proactive agents. Agents can sense, act and interact in order to achieve individual and collective goals. Agents are grouped into coalitions and assigned to specific task (operational level). Agents collaborate with each others based on multi-dimensional trust and reputation model to achieve global goal. Tasks are assigned to agents' teams using optimization technique based on neural network. Then, agents' teams bid for resources to cope with unexpected resource allocation situations.

**Features:**
– System is based on decentralized architecture.
– Machine learning algorithms and control applied at the strategic, operational and tactical levels.
– Improved situational awareness via sensors network.
– System divides global goal (strategic level) into sub-goals (tactical level) represented as sub-graphs, which are executed through agent actions (operational level).
– System adapts to environmental changes via sensors network.
– System involves adaptive on-line decision making algorithms.
– Minimal agent communication.
– System involves auctions to make agents' coalition.
– System involves data fusion techniques.
– System adopts inference and prediction to predict agent future events.
– Flexibility and reusability of agents.
– Crisis real-time response support system.

**Limitations:**
– System architecture does not support fault-tolerance.
– System does not support plug-and-play capability.
– System lacks calibration tools of agent behavior.
– System lacks powerful user interface.
– Agent presents limited configurability in terms of decision, motion and health models.

## 4.4 RoboCup Rescue (ResQ Freiburg Project) [13], [18]:
RoboCup Rescue is a user-centric large-scale simulation for urban-search and rescue. The main design goal of RoboCup Rescue is to enable rescue teams to effectively cooperate despite sensing and communication limitations.

**Architecture:** RoboCup Rescue is a multi-agent simulation and modeling System. System components include simulation engine, knowledge base of agent relations, debugging tools, and data mining software for task evaluation. The RoboCup Simulation league is divided into two subunits, (i) Agent Simulation, and (ii) Virtual Robots. Agent simulation platform currently runs a kernel which connects Traffic simulator, Fire simulator, and Civilian simulator. While, Virtual robot is based on Urban Search And Rescue Simulation (USARSim). USARSim enables users to simulate multiple agents whose capabilities closely mirror those of real robots.

**Methodology (ResQ Freiburg Project):** The basic task of agents in RoboCup Rescue is to collect, store, and evaluate information. Then agents choose best actions fitting to the situation to be executed. Agents coordinate with each others to explore crisis space to find civilians. Agents' motion paths are evaluated through methods for hierarchical real-time path planning. Agents predict the life-time of found civilians, and collaborate to optimize rescue actions sequence using genetic algorithms.

**Features:**
– Decentralized control.
– Prediction of hazard material spread (fire spread).
– System adopts high level plan (strategic level plan).
– System adopts space-exploration techniques.
– Improved situational awareness using sensors network.
– Prediction for the civilian's life time via machine learning.
– Minimizing sequence fluctuations via genetic algorithms.
– Dynamic agent role allocation.
– System involves data mining techniques to update task pre-conditions and post-conditions.
– Calibration of agent teams behavior.
– Crisis real-time response support system.

**Limitations:**
– System architecture does not support fault-tolerance.
– System does not support plug-and-play capability.
– Plans are not adapted to situation changes.
– Agents lack reusability.

## 4.5 FireGrid [3], [7], [22]:
FireGrid is a task-centric collaborative community to pursue research for developing real-time response systems using the Grid. FireGrid addresses response process in the built environment, where sensor grids in large-scale buildings are linked to super-real time grid-based simulations.

**Architecture:** FireGrid is based on task-centric I-X agent architecture. FireGrid integrates several core technologies such as: (i) fire and structural models, (ii) wireless sensors in extreme conditions with adaptive routing algorithms, (iii) grid computing which involves sensor-guided computations, and mining of data streams for key events, and finally (v) command-and-control using knowledge-based Hierarchical Task Network (HTN) planning techniques with user guidance.

**Methodology:** All system components are integrated in the command-and-control (C2) task. The C2 task can be defined as the exercise of authority and direction over available resources towards the accomplishment of some objectives. The C2 process consists of repeated cycles of a number of subtasks similar to tasks adopted in DrillSim agent behavior model. Issues–Nodes–Constraints–Annotations <I-N-C-A> ontology is used to formalize the interactions between the various participating agents towards integrated response-behavior from the perspective of the human controller.

**Features:**
– Grid architecture for distributed computation; OpenMP is used to parallelize sequential codes which is used to predict hazard material spread and required response.
– Improved situational awareness using sensors network.
– Self-Configuring sensors network.
– System supports plug-and-play capability.
– System enables high level plan.
– System supports agent safety and security.
– Crisis real-time response support system

**Limitations:**
– System architecture does not support fault-tolerance.
– Generated plans are not adapted to situation changes.
– Agent presents limited configurability in terms of decision, motion and health models.
– Calibration of system is valid for simple scenarios only.
– System lacks flexibility and reusability of agents.

## 4.6 Analysis of mentioned systems
DrillSim, DEFACTO, FireGrid, and ALADDIN systems are limited to study fire evacuation crises. While, RoboCup Rescue is focusing on urban search and rescue operations. DrillSim and ALADDIN systems are of noticeable participation for matching domain requirements; in which DrillSim is designed to study information technologies metrics in crisis response, and ALADDIN is designed to study different agent architectures in response operations. Current systems development doesn't follow any standards in spite of existing standards waiting to be adopted in response systems [8]. In addition, current systems had focused on roughly supporting response activities with small interest on improving the effectiveness of response operations. System development should take in consideration domain requirements to increase response effectiveness.

## 5. CONCLUSION AND FUTURE WORK

The development of crisis response systems should be directed by crisis response domain requirements. Response requirements are identified by several distinctive characteristics and factors relevant to managing them such as: (i) crisis overwhelms available resources, (ii) crisis requires an immediate response, (iii) crisis event is unpredictable, (v) uncertainty and incompleteness of information and resources, (vi) special need for information, (iv) and required response to prevent secondary crisis. Systems design should include tools for decentralized control, coordination of actors, resource management, data fusion, distributed decision making, adaptation to environment changes, and learning from experience. Our analysis refines in details these required response requirements and mapping of response requirements to system implementation.

We are focusing on our future work on the design of self-defensible and adaptable system model for crisis response. Model development will be directed by domain requirements and will be based on the metaphor of Artificial Immunity System (AIS). Proposed system model will be evaluated by enhancing healthcare response effectiveness and decision support to pandemic diseases such as bird flu.